\documentclass{emulateapj}
\usepackage{times}
\usepackage{amsmath}

\usepackage[pdftex]{color}

\definecolor{myred}{rgb}{0.6,0,0}
\definecolor{myblue}{rgb}{0,0.2,0.4}

\def\la{\mathrel{\mathpalette\fun <}}
\def\ga{\mathrel{\mathpalette\fun >}}

\def\fun#1#2{\lower0.837ex\vbox{\baselineskip0ex\lineskip0.209ex
  \ialign{$\mathsurround=0ex#1\hfil##\hfil$\crcr#2\crcr\sim\crcr}}}

\def\msun{M_\odot}

\def\sles{\lower2pt\hbox{$\buildrel {\scriptstyle <}
   \over {\scriptstyle\sim}$}}

\def\sgreat{\lower2pt\hbox{$\buildrel {\scriptstyle >}
   \over {\scriptstyle\sim}$}}

\def\la{\mathrel{\mathpalette\fun <}}
\def\ga{\mathrel{\mathpalette\fun >}}

\begin{document}

%\begin{landscape}

%\end{landscape}

%\begin{sideways}

%\end{landscape}
%\begin{pdflscape}
%\end{pdflscape}

\title{Galaxy Strategy for LIGO-Virgo Gravitational Wave Counterpart Searches}
\shortauthors{Gehrels et al.}
\author{ 
           Neil~Gehrels\altaffilmark{1}, 
        John~K.~Cannizzo\altaffilmark{2,3},
          Jonah~Kanner\altaffilmark{4},
          Mansi~M.~Kasliwal\altaffilmark{5},
         Samaya~Nissanke \altaffilmark{6},
            Leo~P.~Singer\altaffilmark{1,7}
}
\altaffiltext{1}{NASA Goddard Space Flight Center,
             Mail Code 661, Greenbelt, MD 20771, USA}
\altaffiltext{2}{CRESST and Astroparticle Physics
      Laboratory, NASA/GSFC, Greenbelt, MD 20771, USA} 
\altaffiltext{3}{Department of Physics, University of
       Maryland, Baltimore County, 1000 Hilltop Circle,
           Baltimore, MD 21250, USA}
\altaffiltext{4}{LIGO, California Institute of Technology, 
                            Pasadena, CA 91125, USA} 
\altaffiltext{5}{Observatories of the Carnegie 
                   Institution for Science, 
         813 Santa Barbara St, Pasadena, CA 91101, USA}
\altaffiltext{6}{Institute of Mathematics, Astrophysics and Particle Physics,
                  Radboud University,
                    Heyendaalseweg 135,
                      6525 AJ Nijmegen, The Netherlands}
\altaffiltext{7}{NASA Postdoctoral Program}

\begin{abstract}
In this work we continue a line of inquiry begun in Kanner et al. which detailed a strategy for utilizing telescopes with narrow
fields of view, such as the \emph{Swift} X-ray Telescope (XRT), to localize gravity wave (GW) triggers from LIGO/Virgo.
If one considers the brightest galaxies that produce $\sim$50\% of the light, then the number of galaxies inside typical GW error boxes
will be several tens. We have found that this result applies 
       both in the early years of Advanced LIGO  when the range is small and the error boxes large,
 and in the later years when the error boxes will be small and the range large.
      This strategy has the beneficial property of reducing the
number of telescope pointings by a factor 10 to 100 compared with tiling the entire error box.
  Additional galaxy count reduction will come from a GW rapid distance estimate which
will restrict the radial slice in search volume. Combining the bright galaxy strategy with a convolution based on anticipated GW
localizations, we find that the searches can be restricted to 
       about $18\pm5 $ galaxies for 2015, about $23\pm4$ for 2017, and about $11\pm2$ for 2020.
       This assumes a distance localization at or near the putative NS-NS merger range for each target year,
               and these totals are integrated out to the range. Integrating out to the horizon  would
                      roughly double the totals. 
         For nearer localizations the totals would decrease.
      The galaxy strategy we present in this work will enable numerous 
   sensitive optical and X-ray telescopes with small fields of view to
participate meaningfully in searches wherein the prospects 
         for rapidly fading afterglow place a premium on a fast response time.
\end{abstract}


\section{Introduction}

    The advent of Advanced LIGO (aLIGO -- Aasi et al. 2015) and 
  Advanced Virgo (AdV    -- Acernese et al. 2015) heralds the dawn of a new age of
  discovery in which gravitational wave (GW)
  detections will supplement traditional electromagnetic (EM)
  detections, such
   as those by \emph{Swift}
            (Gehrels et al. 2004). 
      We refer to aLIGO/AdV 
         in combination   
          as LVC.
            These GW observatories will begin
  operating soon at a fraction of design capability,
 and within a few years should be able to detect
   neutron star-neutron star (NS-NS) mergers out to $\sim$450 Mpc
 and black hole-neutron star (BH-NS) mergers out to $\sim$900 Mpc
   (Aasi et al. 2013). These distances are referred to
as the ``horizon'' for these events.

It is important to distinguish between ``horizon'' and ``range''.
   The ``inspiral range'', the most commonly cited
      figure-of-merit as regards LVC sensitivity,
              is defined as
   the radius of a sphere whose volume equals the 
              sensitivity volume
      within which a signal-to-noise $\ge8$ detection
    is achieved
          for a $1.4-1.4\msun$ NS-NS merger,
          averaged over all sky locations and binary inclinations.
     The antenna projection functions and associated averagings
       were discussed by Finn \& Chernoff (1993) and Finn (1996).
   The inspiral range
          is not a hard upper limit, as face-on binary orbit mergers produce
    stronger signals
   (Dalal et al. 2006;
      Maggiore 2007;
        Schutz 2011;
          Nissanke et al. 2013).
     The maximum theoretical detection distance,
          the horizon, is 2.26 times the range 
   (Finn \& Chernoff 1993; see also Abadie et al. 2010).
    A simple calculation shows that
  in    a general population
  with binaries of
       random inclinations and positions over the whole sky, the fraction of
      aLIGO detections one expects to pick up from beyond the nominal inspiral range
      is about half of the total
             (Nissanke et al. 2010, 2013; Singer 2015).

  High signal-to-noise (SNR) detections will permit
    important physical parameters
  to be measured that are 
 not easily accessible through
traditional means, for instance the masses and spins
    of the merger components and the luminosity distance $D_L$
to the merger.   
    A joint EM/GW detection would provide powerful constraints
  on the merger physics.
  Various groups have begun to examine prospects for finding
 an EM counterpart to a GW trigger
           (Nuttall \& Sutton 2010;
    Hanna, Mandel, \& Vousden 2014;
            Ghosh \& Nelemans 2014;
      Fan, Messenger, \& Heng 2015).
                    Evans et al. (2015)
   consider strategies for X-ray observations using \emph{Swift}/XRT.

In Section 2 we present an overview of short GRBs, including putative
GW rates for NS-NS mergers, and a discussion of ``kilonova'' emission  
             accompanying the decay of radionuclides.  
    In Section 3 we describe a galaxy catalog -- the Census of the Local Universe --
   and quantify its completeness out to distances relevant for aLIGO.
In Section 4 we estimate galaxy totals concomitant with putative
aLIGO error volumes for the next 5 yr, if one restricts attention to
 bright  galaxies.  
   In Section 5
    we  discuss  potential EM observations in the  optical, X-ray, and radio
      which might 
        follow  GW detection, and in Section 6 we sum up.

\section{aLIGO-AdV (LVC)  Observations}

\subsection{GW radiation}

  Astrophysical sources that are powerful in EM detectors
      are always weak in GW if the underlying physical process is nearly spherically
  symmetric. Supernovae are among the brightest EM transients
    but might produce peak GW strain amplitudes
    of $|h|D \la 10$ cm, where $h$ is the dimensionless strain and $D$ the distance to source,
   and so be observable in aLIGO only 
to a few kpc (Ott et al. 2013, see their Fig. 14).

  In fact the lowest order contributions to GW radiation
  arise from a changing mass quadrupole moment. As such, binary mergers
    of compact objects give by far the strongest signals,
               yielding peak GW strain amplitudes $|h|D \sim$1 km.
  Belczynski, Kalogera, \& Bulik
    (2002) utilize a population synthesis code
 to delineate a range of rates 
  for different types of binary mergers.
     They calibrate NS-NS
       merger rates
          using known galactic binary pulsars.
        For aLIGO they infer 
       $1-400$ yr$^{-1}$ for NS-NS,
       $9-400$ yr$^{-1}$ for BH-NS,
   and
       $0-8000$ yr$^{-1}$ for BH-BH. 
  Abadie et al. (2010) incorporate 
     the results of the Belczynski et al study
    and  estimate
      an LVC          NS-NS merger rate $\sim$40 yr$^{-1}$,
      with a range $\sim$0.4$-$400 yr$^{-1}$.
   A conservative lower limit to the NS-NS rate for a putative aLIGO inspiral range
     of 200 Mpc is $\sim$3 yr$^{-1}$ (Phinney 1991;
       for a more up-to-date study, 
          see O'Shaughnessy \& Kim 2010). 
 More recently Dominik et al. (2013) calculate local (i.e., $z\simeq0$)
   rates of $\sim$$10^2$ Gpc$^{-3}$ yr$^{-1}$ for NS-NS
   mergers
  and   $\sim$$10  $ Gpc$^{-3}$ yr$^{-1}$ for BH-NS mergers. 
        Given a putative LVC sensitivity volume $\sim0.03$  Gpc$^3$
   relevant for a 200 Mpc range,
         this translates in LVC detection rates 
            of $\sim3$ yr$^{-1}$  
         and $\sim0.3$ yr$^{-1}$, respectively, similar to Phinney (1991).
                      As indicated earlier, 
             taking into account 
    the more realistic totals achieved in mapping from range to horizon would
        roughly double these estimates.
   In this work we restrict our attention to NS-NS mergers.

\subsection{GRBs and beamed EM radiation}

   GRBs come in two flavors:  long ($>2$ s) and short ($<2$ s)  (Kouveliotou et al. 1993).
     Long GRBs (lGRBs) are thought to arise from the explosion of a massive
star - nearby lGRBs have associated supernovae. 
    More interesting for aLIGO are short GRBs (sGRBs) which are 
thought to be due to NS-NS mergers for which we are placed
  along the binary rotation axis (Eichler et al. 1989;
     Narayan, Paczy\'nski, \& Piran 1992). 
    The fact that lGRBs are mostly        spherically or axially
  symmetric makes them less interesting than sGRBs
 vis a vis GW detections.

   The  observed redshift range is from
  about 0.2 to 2 for short GRBs (sGRBs),
  with a mean of about 0.4.
  The XRT commonly observes sGRBs up to $z\simeq0.5$
   but in 10.5 yr of operation \emph{Swift} has not seen one 
      with a measured redshift $z < 0.1$, 
 i.e., the NS-NS merger aLIGO target GW horizon $\sim$400 Mpc,
     suggesting that such events may be quite rare.

   Estimates for jet beaming
   are 
      $\theta_{\rm j}\sim5^{\circ}$ for lGRBs
          and
     $\theta_{\rm j}\sim5-15^{\circ}$ for sGRBs
 (Burrows et al. 2006, Grupe et al. 2006, Fong et al. 2012).
  Beaming angles for sGRBs
  are still highly uncertain.
  The
  beaming factors
 $f_b = 1-\cos\theta_{\rm j}\simeq \theta_{\rm j}^2/2$
 are  roughly $1/300$ for lGRBs and $1/30$ for sGRBs.
       Based on the observed rate of sGRBs by 
  \emph{Swift}, Coward et al. (2012)
      estimate a   LVC        detection rate of 
  $\sim$$3-30$ yr$^{-1}$ for  $\theta_{\rm j}\simeq 15^{\circ}$. 
  Chen \& Holz (2013) claim $3-7$  yr$^{-1}$
   for GRB GW+EM detections.
   Kelley et al. (2013) estimate the rate of \emph{Swift} or \emph{Fermi}
     observations joint with LVC detections to be $\sim$0.07 yr$^{-1}$.
  Siellez, Bo\"er, \& Gendre (2014) consider current and future 
  high energy missions and estimate a rate of simultaneous GW+EM detections
     of $\sim$$0.1-4$ yr$^{-1}$ in the  LVC     era.
  Wanderman \& Piran (2015)  estimate a co-detection
  rate
    LVC$+$\emph{Fermi} of  $0.1-1$ yr$^{-1}$ and
    LVC$+$\emph{Swift} of $0.02-0.14$ yr$^{-1}$.

\subsection{Kilonova emission}

  Various groups have explored the
   supernova-like
  transient powered by radioactive decay
            of the spray of material
    $\sim$$10^{-6}-10^{-1}\msun$
           ejected from the NS
  (Eichler et al. 1989;
   Li \& Paczy\'nski 1998;
   Metzger et al. 2010;
   Metzger \& Berger 2012).
    The resultant  ``kilonova''  (dimmer than a supernova and brighter than a nova)
         would
         produce
    relatively isotropic optical/NIR
   emission after a NS-NS/BH-NS merger.
    While SNIa  light curves are powered
 primarily by decay of $^{56}$Ni,
    the ejecta from a disrupted NS is neutron
 rich and yields little Ni.
        Much heavier radioactive
elements form via rapid neutron capture
  ($r-$process)
  nucleosynthesis   following the decompression
    of the ejecta from nuclear densities.
  These
newly synthesized elements undergo nuclear
 fission, $\alpha$ and $\beta$
decays on much longer time-scales. 
The resulting energy release can power detectable thermal emission
once the ejecta expands sufficiently that photons can escape.
  Recent general relativistic NS-NS merger simulations 
   (Bauswein, Goriely, \& Janka 2013)
      indicate that a small fraction of the ejecta, $\sim$$10^{-4}\msun$, or a few percent, 
   expands rapidly enough for most neutrons to escape capture.
    The $\beta$-decay of these free neutrons in the outermost
ejecta powers a precursor  to the main kilonova emission
   peaking on a timescale of hours after NS-NS merger (Metzger et al. 2015).
  For $D\simeq200$ Mpc this emission peaks in the $U-$band  ($\sim$$365$ nm)
  at $m_U\simeq 22$ (Metzger et al. 2015).

    Kasen et al. (2013)  argue that the
  opacity of the  expanding $r-$process material is
dominated by bound-bound transitions from those ions
    with the most complex valence electron structure,
   i.e.,
the lanthanides
 (Tanaka \& Hotokezaka 2013;
    Grossman et al. 2014;
     Wanajo et al. 2014). 
         They compute
    atomic structure models for a few representative ions
         in order to calculate the radiative transition
rates for tens of millions of lines, and
     find that  resulting $r-$process
   opacities are orders of magnitude larger than that of
ordinary (e.g., iron-rich) supernova ejecta.
The resultant light curves should be longer, dimmer,
   and redder than  previously thought.
  The spectra
   have broad absorption features and peak in the IR
  ($\sim$1 $\mu$m).
  Kasen et al. (2015) combine two-dimensional hydrodynamical
  disk models with wavelength-dependent radiative transfer calculations
 to generate model light curves and spectra.
    They discern two components to the kilonova
light curve, a blue optical transient ($\sim$2 d) arising from the
  outer lanthanide-free ejecta and an IR transient
  ($\sim$10 d) coming from the inner, lanthanide line-blanketed region.
   There had been an earlier  suggestion 
   of these two components 
   in work by Barnes \& Kasen (2013)
   using a less sophisticated model.

   Specific predictions for kilonova light curves  are dependent
     on uncertainties such as the
   type of ejecta (dynamical and/or disk outflows),
 ejecta masses, 
             and velocities.
       Recent time dependent calculations  find ejecta
   masses in the range $\sim$$10^{-3}$ -- $10^{-1}\msun$
   and velocities $\sim$0.1--0.3$c$
  (e.g., 
        Foucart et al. 2011;
        East \& Pretorius 2012;
         Piran et al. 2013;
    Kyutoku et al. 2013;
  Hotokezaka et al. 2013;
 Fern{\'a}ndez \& Metzger 2013;
   Sekiguchi et al. 2015; 
     Foucart et al. 2015;
  Fern{\'a}ndez et al. 2015).
 Assuming iron-rich supernova ejecta,
     Metzger \& Berger (2012)
          predict peak
    optical  luminosities $\sim$$10^{41}$--$10^{42}$ erg s$^{-1}$
  and concomitant $M_R$ values       $-14$ to $-17$.
   For $M_{\rm eject}=10^{-2}\msun$ and $v_{\rm ejecta} = 0.1c$,
   Barnes \& Kasen (2013) calculate 
          a peak absolute magnitude in the near-IR ($\lambda\simeq 1.7$$\mu$m)
              of   $M_H =  -15.5$,
  in good agreement with Tanaka \& Hotokezaka (2013), 
      Tanaka et al. (2014), and  Grossman et al. (2014).

GRB 130603B might be the first detected
   kilonova (Tanvir et al. 2013;
        Berger, Fong,
               \& Chornock 2013).
  It  was a short
          GRB at $z=0.356$ with a duration $\sim$0.2 s in the BAT.
If correct, it would confirm that compact-object mergers
  are the
progenitors of short GRBs and also  the sites of significant
production of $r-$process elements.

In addition to the optical/near-IR kilonova emission, 
   one also expects a characteristic signal in the
  radio
 (Nakar \& Piran 2011; Hotokezaka \& Piran 2015)
  as the ejecta first interact internally and
  then externally with the ISM.
   The latter interaction gives rise to a blast wave
with concomitant enhancement of magnetic fields
  and electron acceleration, leading to synchrotron radiation
   and radio emission.
 Three temporal components to the radio band
      have been considered and studied,
(i) early-time anisotropic emission along the relativistic jet
  axis associated with the ultra-relativistic ejecta,
 (ii) mildly relativistic, quasi-isotropic emission
     accompanying cocoon-breakout, leading to potential radio
   flares for off-axis observers, and 
(iii) late-time sub-relativistic dynamical ejecta producing radio flares
     on time scales of years. The latter emission should be 
  nearly isotropic, and provide standard calorimetry on the
global energetics of the initial explosive event, just
   as has been the case for long GRBs.

  \section{Census of the Local Universe catalog}

Any given galaxy catalog is generally not optimal for GW follow-up studies.
  Consider two extremes:
    The 
   Two Micron All Sky Survey (2MASS)
  survey (Skrutskie et al. 2006)
        has good coverage in both the northern and southern skies,
      but does not go very deep (Huchra et al. 2012).
     The Millenium Galaxy Catalog,
  comprising spectroscopic redshifts of galaxies
  from 2dF or SDSS, is deep, but covers only a small slice 
     along the celestial equator (Driver et al. 2005).
   An attempt to overcome these limitations 
   led to the 
   Gravitational Wave Galaxy Catalogue $=$ GWGC
  (White, Daw, \& Dhillon 2011).
    Its only limitation
        for our current study is that it
         does not extend beyond 100 Mpc.
      One of us (Kasliwal et al. 2015, in preparation) 
      has amassed a catalog based on the union of
several existing catalogs -- 
      the Census of the Local Universe (CLU)  --
                 which is suitable for GW+EM follow-up studies.
  As we shall show, for bright galaxies
    the CLU  is complete out to the anticipated aLIGO GW inspiral range
  for NS-NS mergers
   up to 2020.

\begin{figure}[h!]
\begin{centering}
\includegraphics[width=4.20truein]{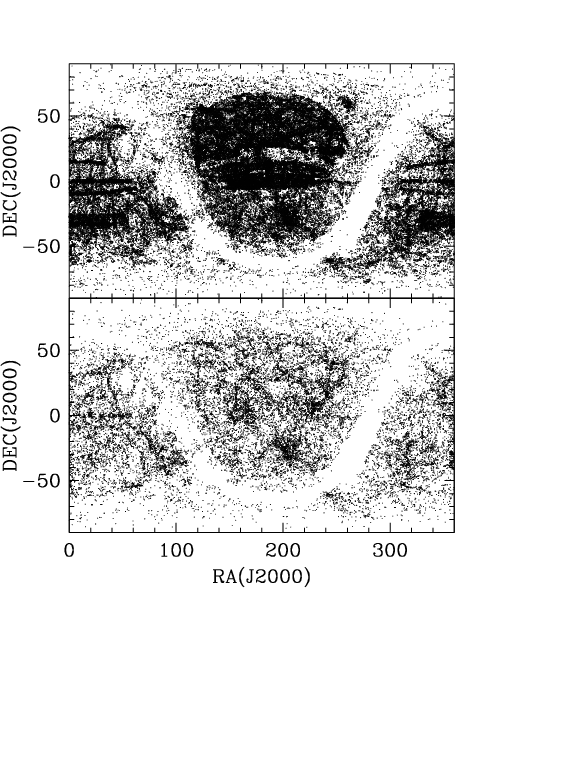}
\vskip -1.0truein
\caption{
  Sky maps of the CLU catalog. 
  Shown are all 144214 galaxies ({\it upper panel}),
  and only the 27559 galaxies for which $L > L_B^*$
        ({\it lower panel}).     The swath of
    incompleteness in both panels represents the 
 galactic plane, which is excluded in many surveys.
    }
\end{centering}
 \end{figure}

In order to show completeness, we must adopt a model for
   galaxy number density in the local universe.
The Schechter luminosity function (Schechter 1976)
  provides
  a useful description
  of the space density of
 galaxies as a function of their
   luminosity,
  $\rho_{\rm gal}(x) dx = \phi^* x^a e^{-x} dx$,
  where $x=L/L^*$ and
   $L^*$ is a characteristic galaxy luminosity
  where the power-law form of 
  the function truncates.   It has proven to be applicable
     over up to 10 magnitudes
       in deep surveys (e.g., Bonne et al. 2015).
 The CLU catalog is amassed from many different surveys.
    One of the CLU data columns, $b_{\rm tc}$, consists of apparent $B-$magnitudes
       $m_B$  for entries where they are available, and pseudo-$m_B$ values
        for sources from
   other bands, for instance 2MASS.   
      Hence for this work
  $\phi^*$, $L$,  $L^*  \rightarrow {\phi_B}^*$, $ L_B$, $ {L_B}^*$.
    Physically, $L_B^*$ represents the turn-over in the 
 distribution between a power-law              for low $x$
  and an exponential         for high $x$.

  We adopt the following
  values derived from $B-$band measurements
  of nearby field galaxies:
      $\phi^* = (1.6\pm0.3)\times 10^{-2} h^3$ Mpc$^{-3}$,
        $a=-1.07\pm0.07$,
          $L_B^* = (1.2\pm0.1) \times 10^{10} h^{-2} L_{B, \sun}$,
  with a corresponding
  $M_B^* = -19.7 \pm 0.1 + 5 \log_{\rm 10} h = -20.47$
    (e.g., Norberg et al. 2002;
    Liske et al. 2003;
    Gonz\'alez et al. 2006,
   and references therein). 
          By comparison, for the 
   Milky Way galaxy  $M_B = -20.42$.
       This is based on a Milky Way $B-$band
            luminosity $2.3\times 10^{10} L_{B, \odot}$ (Carroll \& Ostlie 1996),
  where $L_{B, \odot}$ is the solar $B-$band luminosity,
     and an absolute solar $B$ magnitude of 5.48 (Allen 1973).
  We take $h=0.7$ based on the latest weighted overlap
   between the \emph{Planck} results and the rest of astronomy
  (Ade et al.
  2014).

\begin{figure*}
\begin{centering}
\includegraphics[width=7.20truein]{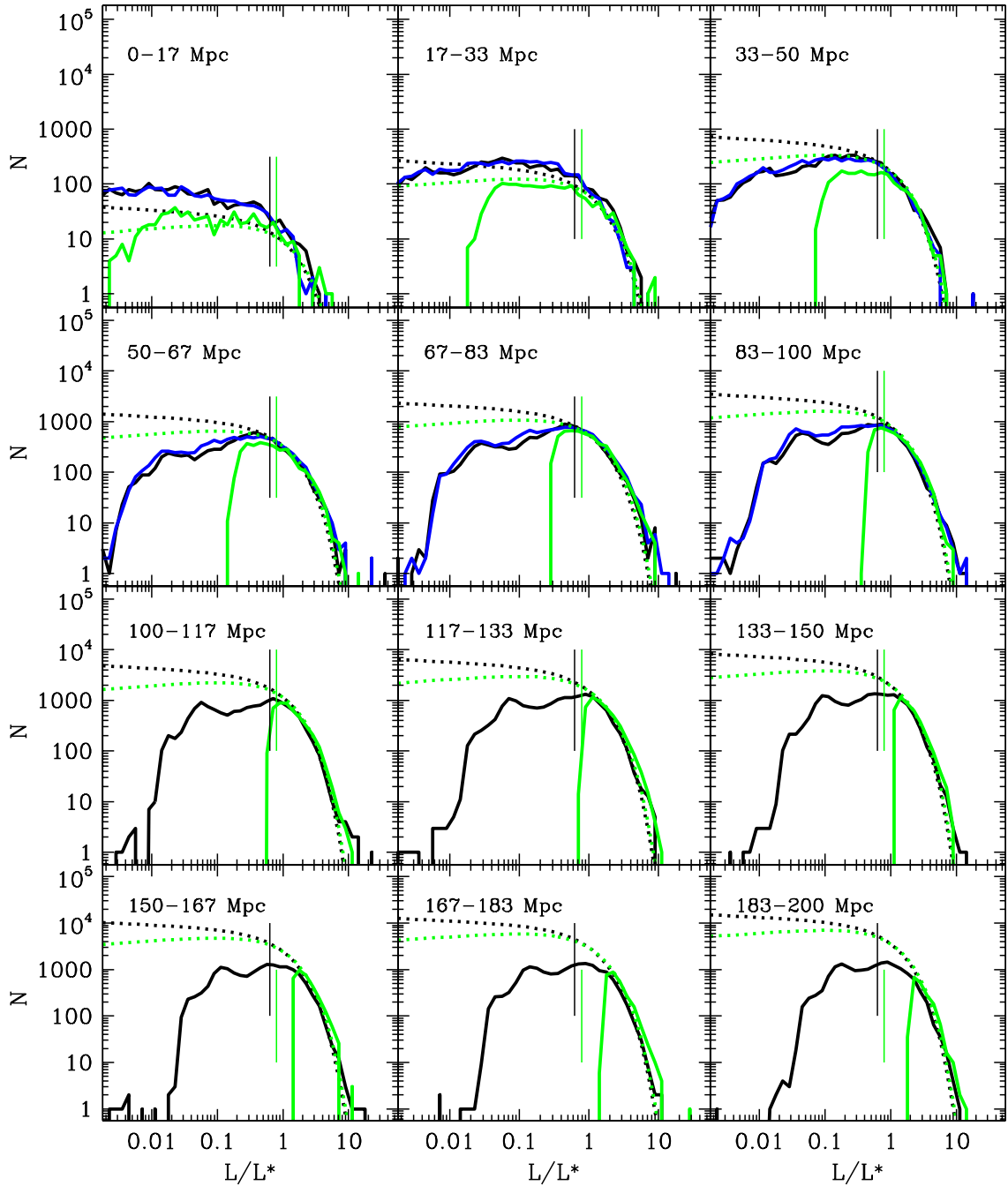}
\vskip -0.5truein
\caption{ A comparison of the completeness
  measures of the CLU catalog ({\it black}) 
        with GWGC ({\it blue})  and 
      the 2MASS redshift survey ({\it green})
  is given by showing frequency histogram distributions
   for 12 distance slices ({\it solid}) versus $x=L/L^*$,
      using bins of width 0.1 dex in $x$.  
       For CLU and GWGC $x=L_B/{L_B}^*$,
           whereas for 2MASS $x=L_K/{L_K}^*$.
                   Shown also is the Schechter function 
     $\phi^*  x^a  e^{-x}  \Delta V  dx $
     ({\it dotted}), 
          where the volume
  element $\Delta V$ is that for the given distance slice.
       For the 2MASS data,  which are in the $K-$band, $\alpha=-0.9$
   and 
             $x_{1/2}=0.790$. 
      The vertical line segments in each panel
                   indicate $x=x_{1/2}$. 
  %   [{\bf Note to editor: we request a full page for this figure.}]
}
\end{centering}
 \end{figure*}

Integrating over luminosity
             gives  integrated number density 
                    $\rho_{\rm 0,gal} = 
   \int_{x_1}^{\infty} \rho_{\rm gal}(x) dx$. 
      Although  
           $\rho_{\rm 0,gal} \rightarrow \infty$
  as $x_1 \rightarrow 0$ for $a < -1$,
 the integrated luminosity density diverges only for  $a<-2$. 
   One has
\begin{equation}
       \int_{x_1}^{\infty} 
   \phi^*  L^*
          x^{a+1} \exp(-x) dx
    = 
   \phi^*
    L^*
    \Gamma(a+2, x_1),
\end{equation}
  where $\Gamma$ is the incomplete gamma function.
    For $a=-1$
  the total luminosity density is
  $\phi^* L^* \Gamma(2+a)
 = \phi^* L^* =   1.9\times 10^8 h L_B(\odot)$ Mpc$^{-3}$.
     Dividing      by a Milky Way $B-$band
            luminosity    
  yields a density $\sim$$6\times 10^{-3}$ MWE Mpc$^{-3}$,
  where MWE $=$ Milky Way equivalent galaxy.
    For $a=-1$,  half of the luminosity density is contributed by galaxies
with $x_{1/2} > 0.693$. For the power law of interest in this study,
  $a=-1.07$, the cutoff lies at  $x_{1/2} > 0.626$,
 or  $M_{B \ 1/2} = -19.97$. This corresponds to $\sim$0.66 of the Milky Way luminosity.
  To arrive at this  $x_{1/2}$ value
            we used the fact that
     $\int_{x_1}^{\infty} x^{a+1} e^{-x} dx =  1.04559$ for $x_1=0$ and $a=-1.07$,
   and half this value $ 1.04559/2$ is achieved for $x_1 = 0.626$.

Figure 1 presents sky maps of the CLU catalog, showing all the galaxies,
             and also those 
         for which  $x>1$,   where $x = L_B / {L_B}^*$.
  The dark strips evident in the top panel indicate individual
deep surveys which make up the CLU. Restricting the sample to
intrinsically bright galaxies cleans it up considerably
   and reveals large scale structure.

      Figure 2 provides a measure of the completeness
 as gauged by the Schechter function.
   We compare the CLU catalog with the GWGC and 2MASS redshift survey
  out to 200 Mpc using 12 radial slices.
      Within each slice we bin the data in  $x= L/L^*$
    and compare  with 
   the  Schechter function  
   weighted by
  the volume of the given slice. 
     For CLU and GWGC $x=L_B/{L_B}^*$ whereas for 2MASS $x=L_K/{L_K}^*$.
 Since GWGC ends at 100 Mpc there are no data in the more distant bins.
  For all three catalogs
    there is a progressive loss of fainter galaxies
   with distance relative to
  Schechter.
      For all radii the CLU follows 
   the Schechter function  
   for  $x \ga x_{1/2}$.
      In some of the panels in Figure 2 one sees 
      spikes at $x\simeq 10$.  These may be due to  
      misidentified local objects
              spuriously placed at larger distances.

\begin{figure}[h!]
\begin{centering}
 \includegraphics[width=4.05truein]{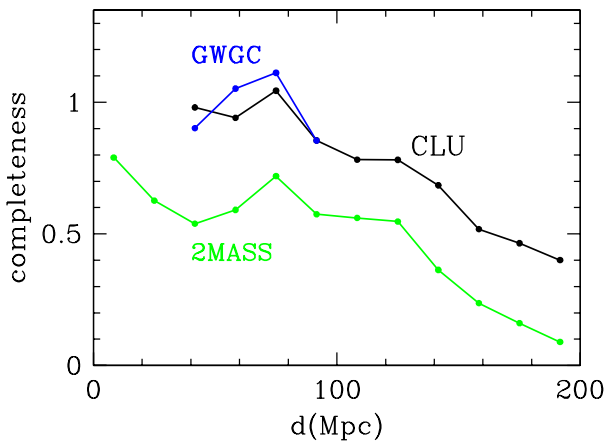}
\vskip -1.95truein
\caption{Completeness  
        relative to the Schechter function
     for $x > x_{1/2}$  for CLU ({\it black}), GWGC ({\it blue}), 
     and 2MASS ({\it green})
    using
 the data presented in Figure 2.
     }
\end{centering}
 \end{figure}

     In this paper we propose to utilize the brighter galaxies in each
     bin, which tends to mitigate the effects of incompleteness
     which are brought about by a progressive loss of fainter
     galaxies with increasing distance.  
The CLU catalog is fairly complete relative to Schechter
    above 
 $ x_{1/2}$.
   This is shown in Figure 3,
    where we compare with GWGC and 2MASS.
              In this work we will argue for including only the brighter galaxies
    in GW+EM follow-up studies. As noted earlier, this implies cutting galaxies
      below $x_{1/2}$.  
 For the three fiducial distances of interest in the next section, 60 Mpc, 120 Mpc, and 180 Mpc,
   the CLU catalog is complete above this cut line 
      at a level  $\sim$100\%, $\sim$80\%, and $\sim$40\%,
     respectively.  Since these three distances are based on the NS-NS inspiral ranges
   for 2015, 2017, and 2020,
        an additional factor which must be included for the third year 
 stems from the fact that the CLU catalog ends at 200 Mpc, namely the fact that 
     about half the detections would be expected to arise from between the range
  and the horizon.   Therefore for the third year the completeness is $\sim$20\%.
      Both Figures 2 and 3  show some regions of ``overcompleteness''
          below 100 Mpc. This is probably due to local over-densities
          such as the
     Virgo Supercluster.

   Our choice of $x_{\rm cut} = x_{1/2}$ is motivated by a trade-off
between including enough galaxies so that we encompass a reasonable fraction
   of putative sGRB host galaxies (e.g., Berger 2014, see his Figure 8)
        on the one hand, and not 
          having incompleteness become too great an issue on the other hand.
 Taking  $x_{\rm cut} = x_{1/2}$ picks out about half of the sGRB host galaxies
   shown in Berger (2014), therefore the true efficiency factors for the three target
     years would be $\sim$50\%, $\sim$40\%, and $\sim$10\%, respectively.

 Another caveat arises from the fact that
       by utilizing $B-$band data we are implicitly taking 
           $B-$band luminosity as a proxy for the compact
      object merger rate. Although this assumption is generally supported
     by sGRB observations (Berger 2014),  
       it may in fact not be a universally good proxy
     (e.g., de Freitas Pacheco et al. 2006;
       O'Shaughnessy et al. 2010; 
      Hanna et al. 2014).

\section{Galaxy Strategy}

   As regards identifying a candidate host galaxy
     for a GW event,
         a galaxy catalog is only half of the picture.
     Of course the starting point is the GW localization.
 Initial localizations in the $\sim$2015 time frame are expected
  to be $\sim$$500$ deg$^2$, decreasing 
    to $\sim$$20$ deg$^2$ by 
      $\sim$2020 (Aasi et al. 2013; Singer et al. 2014;
       Berry et al. 2015).
         In terms of the depth,
the NS-NS inspiral range should increase from $\sim$60 Mpc to $\sim$180 Mpc
over the same time frame. A projected timeline of 
    NS-NS inspiral range versus date
  is given by Aasi et al. (2013).
   One may obtain a good estimate using simple considerations.
    In the CLU catalog there are $N^*=27559$ galaxies
  brighter than ${M_B}^*$, and $N_{1/2}=47438$  
   brighter than  $M_{B \ 1/2}$.
   If one restricts $N_{1/2}$ 
    based on  the limited sky areas and volumes
  relevant for the three target years, i.e., 
 $500$ deg$^2$ $\times$  60 Mpc for 2015, 
 $100$ deg$^2$ $\times$ 120 Mpc for 2017, and
 $20  $ deg$^2$ $\times$ 180 Mpc for 2020, 
   one obtains $N_{\rm gal} = $ 26.0, 36.4, and 18.8, respectively.
           As noted previously, since these distances represent putative NS-NS
      inspiral ranges, for a localization near $\mu$ we must roughly
                      double the totals to take into account going
           from range to horizon if the distance error is large.

  Realistic idealizations for aLIGO 
  localizations have been undertaken by
  several recent groups which attempt to quantify
  the ``volume reductions'' which might realized.
       Nissanke et al.  (2010) 
     use a Markov Chain Monte Carlo (MCMC) technique
    to calculate
    distance measurement errors 
          associated with aLIGO localizations of 
             astrophysical populations of NS-NS and BH-NS binaries,
   considering both isotropically oriented as well as beamed events.
        They take as their starting point a precise sky localization based
    on a coincident EM detection of the same GW event.
       They present  Fisher-matrix-derived linear scalings 
  for $[\Delta D_L/{\rm True} \ D_L]$ for the two populations,
      assuming four GW detectors.
       If the EM emission from the NS-NS merger providing the coincident EM signal is isotropic, 
   they find that, in combination with the precise sky position,
                   the distance to NS-NS binaries
  can be measured with a fractional error of $\sim$$20-60$\%,
 with most events clustered near $\sim$$20-30$\%.
     If the EM emission from the NS-NS merger is beamed, with a $\sim$25$^{\circ}$ opening angle,
   then the error on the distance is reduced by a factor $\sim$2
     and much of the high error tail is eliminated.
       BH-NS events are measured more
         accurately: 
           the distribution of fractional distance
              errors 
     lies in the range
      $\sim$$15-50$\%, with most events clustered near $\sim$$15-25$\%.
    Nissanke, Kasliwal, \& Georgieva (2013) 
   carry out extensive end-to-end simulations,
     looking at GW sky localization, distance errors and volume
  errors using NS-NS and BH-NS mergers. 
  They compare MCMC-derived distance measures with three dimensional (3D) volume measures.
            They outline optimal strategies to prepare 
             for identifying EM counterparts of a GW merger.

   Singer et al. (2015) have 
  created 
  full  3D position reconstructions
  for a large population of 
   simulated early aLIGO 
     NS-NS events.  
   They provide a simple approximation
  for the 3D distance distribution
      and qualitatively describe
  the shapes that emerge.  
  In the present work, as we are more interested in the
   impact of the galaxy catalog we use an even
   simpler description:   we assume that the 
     localization subtends a given solid 
  angle, and is a shell between two constant radii.

\begin{figure}[h!]
\begin{centering}
\includegraphics[width=3.75truein]{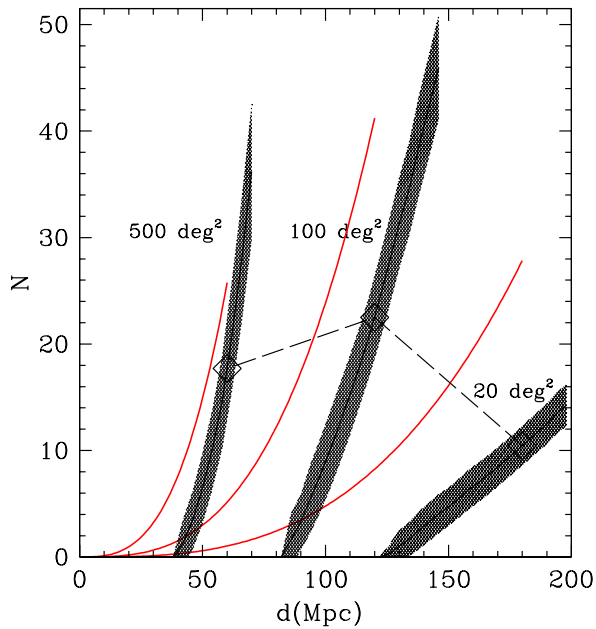}
\vskip -1.0truein
\caption{
The number of CLU galaxies in a
  given size error box versus range 
                                     ({\it black}).
               In this experiment we implicitly assume a localization near the inspiral range
                           for each of the three target years.
                    Diamonds indicate the adopted NS-NS inspiral ranges for the three target years 
                     of this study, $\mu=$ 60, 120, and 180 Mpc, respectively.
   The three curves are representative of
  the increasing localization capability
    of aLIGO+VIRGO  with time, 
    $\Delta\Omega\simeq500$ deg$^2$ (for 2015),
          $100$ deg$^2$ (for 2017), and
            $20$ deg$^2$ (for 2020).
  The CLU incompleteness weighting factors at $r=\mu$
         are 1.0, $1/0.8=1.25$, and $1/0.4             $, respectively. 
   The galaxy count totals indicated by the diamonds are
   $N_{\rm gal}=17.7\pm5$, $22.5\pm4$, and $10.5\pm2$.
  We also present the corresponding  $N_{\rm gal}$ values
      determined directly from the Schechter function ({\it red})
      weighted by the relevant volume $\Delta V = (4/3)\pi \mu^3
  (\Delta\Omega/4\pi)$ (see the end of Section 4).
    }
\label{fig:tiling}   
\end{centering}
 \end{figure}

 Singer et al. (2015) calculate  full 3D aLIGO 
  reconstructions based on BAYESTAR,
   a rapid reconstruction method for BNS mergers,
   and LALInference, a full Bayesian parameter estimation code
  (Singer et al. 2014; Singer 2015; Veitch et al. 2015),
          for the near-future of aLIGO - 2015 and 2016.
   In this work we consider a longer time frame, i.e., up to the full achievement
of aLIGO design sensitivity. 
   In order to calculate galaxy sky counts in error boxes of  
given areas on the sky, we take a simplified approach compared to 
     Nissanke et al. (2009), 
   Nissanke et al. (2013), and Singer (2015).   Namely, rather than
using a realistic 3D reconstruction, for a given line-of-sight (LOS)
   we take a simple top-hat windowing function. 
   For our three putative aLIGO target years -- 
   2015, 2017, and 2020 -- 
    we adopt $\mu= 60$, 120, and 180 Mpc,
       respectively, as fiducial LVC NS-NS inspiral ranges.

          We consider 1000 randomly selected LOSs
   over the sky, and then search the CLU catalog to find galaxies
 within an angular separation that would place them inside an error box
on the sky of  (i)  500 deg$^2$, 
               (ii) 100 deg$^2$,  or
               (iii) 20 deg$^2$  
      for the three cases.
             Although
        the actual sky-projected aLIGO
     localizations will be complicated, the more important factor 
 is simply the
       total sky area involved.
     Only galaxies are considered for which their luminosity places
  them above the 50th percentile mark  $x_{1/2}$. 
    The CLU galaxy count in a radial bin 
        is incremented only if $|(r-\mu)/\mu| < 0.3 $ (Hanna et al. 2014),
      i.e., if the galaxy lies within
    a thick spherical shell with  $\Delta\mu/\mu = 0.6$,
   where the NS-NS inspiral range $\mu = 60$, 120, and 180 Mpc, respectively.  
    This has the effect of essentially doubling the galaxy counts derived 
by truncating the integration at the range $\mu$, thereby mimicking 
the effect of the range-to-horizon mapping (Hanna et al. 2014).
   Thus in this experiment we are implicitly assuming a localization
    near $\mu$ for each target year, and thus in some sense our galaxy count totals
represent upper limits.

     Figure 4 shows the results of this experiment.
     Each point is the mean of the 1000 individual LOS values, 
     and the error bar is the standard deviation.
     Two main points are worth noting. The most obvious is simply
      that larger error boxes yield more galaxies.
   The second is that counts are added only 
               out to roughly  the range for a given year.
The CLU totals out to $r=\mu$ (for a localization in each year
               near $\mu$)
       are 
   $N_{\rm gal} =17.7\pm5$, $22.5\pm4$, and $10.5\pm2$, respectively.
 As noted previously, the CLU is 
     $\sim$100\% complete
    above $x_{1/2}$ at 60 Mpc, 
     $\sim$80\% complete
                    at 120 Mpc, 
and  $\sim$40\% complete
             at 180 Mpc.
   In addition,  the NS-NS range-to-horizon mapping roughly doubles 
        the   totals for localizations  near $r=\mu$ with a large radial uncertainty.

For consistency we may consider galaxy counts derived directly from the Schechter
  function.  The galaxy density above $x_{1/2}$ is $\int_{x_{1/2}}^{\infty}
   \phi^* x^a e^{-x} dx = 2.35\times 10^{-3}$ Mpc$^{-3}$.
   If we then multiply this by a volume $\Delta V = (4/3) \pi \mu^3 (\Delta\Omega/4\pi)$, 
   where, for the three target years ($\mu$, $\Delta\Omega$) $=$ 
    (60 Mpc,  500 deg$^2$),  
   (120 Mpc, 100  deg$^2$),  and
   (180 Mpc, 20   deg$^2$),  respectively, we obtain $N_{\rm gal}=25.8$, 41.3, and 27.8.

\section{Tiling Observations}

  The results given in Figure 4 reveal the dramatic reduction
  in tiling requirements brought about
  by restricting a search methodology to the  brighter       galaxies.
 Were one  simply to tile the entire 2015, 2017,
             and 2020 putative error boxes
 $A= 500$, 100, and 20 deg$^2$ using an EM detector with a field-of-view (FOV)
   of $\delta A = 0.1$ deg$^2$, i.e., $\sim$$19\times19$ arcmin, 
   the number of tilings required 
 would be $A/\delta A  = 5\times 10^3$, 
                                $10^3$, and $2\times 10^2$, respectively.

 Concentrating on selected bright galaxies 
             reduces the tiling effort considerably.
       The optimal size of a ``tile'' is dictated by observations of short
        GRBs - in particular their     locations within their host galaxies.
     Fong, Berger \& Fox (2010) use precise \emph{HST}
 localizations to study the cumulative distribution of projected physical
offsets for short GRBs with sub-arcsecond positions and find them to lie
  within  $\sim$100 kpc of their host galaxy centers.
            For a typical distance of
   interest in this study, 100 Mpc, this corresponds to $\sim$10$^{-3}$ radian or
$\sim$0.057 deg -- a projected area on the sky of $\sim$0.010 deg$^2$
     or $\sim$37 arcmin$^2$.
     This can be covered by an EM detector  
      with a rectangular field-of-view (FOV) of 
           at least $7\times7$  arcmin or 49 arcmin$^2$.
       For such detectors the number of tilings will reduce 
to the number of galaxies as shown in Figure 4,  
  i.e., $\sim$20, $\sim$20, and $\sim$10, 
          respectively. 
         This represents a reduction in requisite tilings 
            by 
             $\sim$$1-2$ orders of magnitude
        over a brute force methodology,
    i.e., simply covering the entire GW error box.

Furthermore, the underlying strategy of  
   focusing attention  only on bright  
  galaxies is strengthened by the observation
                  that short GRBs tend to lie in the larger, brighter galaxies
 (in contrast to long GRBs which lie preferentially in dwarf irregulars).
   Fong et al.  (2010) compare the projected physical offset distribution
between short and long GRBs and find that, when normalized to the sizes of their
  host galaxies, the distributions are indistinguishable. However, the absolute
    length scales differ by
   a factor of $\sim$5 (see also
                            Fong \& Berger 2013).   
  Berger (2014, see his Figure 8) plots all known sGRB host galaxy
        $x=L_B/{L_B}^*$ values;
    they span a range $0.1 \la x \la 2 $.
   Our adopted cut value in this study $x_{1/2}\simeq 0.6$ is roughly at the median
    of the observed distribution, therefore our bright galaxy strategy would pick
up about half the sGRBs shown in Berger's sample. 
   Therefore the total effective completeness would be reduced
      by another factor $\sim$2 below that given in the previous section,
         which only considered completeness for $x > x_{1/2}$. 
The trade-off against
lowering our cut value so as to include a greater fraction
    of putative sGRB host galaxies, 
            i.e., using  for example $x_{1/3}$ or $x_{1/4}$, 
   is two-fold: (i) the number of galaxies would increase rapidly and 
   (ii) incompleteness would become a more severe issue.

\subsection{Optical}

  Within an area of several square degrees on the sky
     there are many more optical transients than in other wavelengths,
         and therefore more opportunities 
              for false positives 
   (Kulkarni \& Kasliwal 2009;
    Drout et al. 2014;
   Tanaka et al. 2014;
   Cowperthwaite \& Berger 2015;
    Singer et al. 2015).
   Most of this activity stems from variable sources within our own galaxy.
            If we restrict our attention to bright galaxies, 
                    false positives could still arise from powerful sources within the
   target galaxy  such as supernovae (SNe).
     However, SNe are rare 
    and also have longer timescales than EM
      counterparts to NS binary mergers. 
    More generally,  Nissanke et al. (2013) and 
         Kasliwal \& Nissanke (2014) show that false positives 
  can be due to both foreground stars (e.g., flare stars and cataclysmic variables)
         and background galaxies
(e.g., supernovae and AGN). 
       Chance associations of such transients with a host galaxy 
       location would be problematic to exclude based solely on photometry,
       and would probably necessitate multiwavelength observations.
          An      advantage with the optical (and near-IR) is that 
               the emission of interest is due to the kilonova,
    which is quasi-isotropic, whereas X-ray emission, for instance,
          would be beamed.

A variety of large optical telescopes that have been active
   in GRB follow-up have large FOVs which would make them amenable
     to GW+EM tiling observations:
  GTC/OSIRIS (10.4m aperture  --  $7.8\times8.5$  arcmin FOV),
  Keck/LRIS  (10m          --  $6  \times8  $  arcmin),
  LBT/LBC    (8.4m      --   $23\times23$  arcmin), and
   VLT/VIMOS (8.2m   --   $14\times14$  arcmin).
  These instruments could all cover the Fong et al. (2010) short GRB projected
    area in one tile.

\subsection{X-ray}

 The transient sky is less chaotic in X-rays than in the optical.  Therefore individual transients
         stand out more.
   However, for short GRBs only the beamed events will be detected in X-rays, 
        which reduces the detection 
   chances by the reciprocal of the beaming factor $f_b$ of a short GRB,
    which is highly uncertain, 
            $f_b^{-1} \simeq 10-100$.
  The primary instrument  of relevance for X-ray follow-up
    is the 
       \emph{Swift}/XRT (Burrows et al. 2005),
        a focusing X-ray telescope with a 110 cm$^2$ effective area, an
  18 arcsec resolution (one-half power diameter) in the $0.2-10$ keV band, 
   and a field-of-view (FOV) of $23.6\times23.6 $ arcmin, or $\sim$0.15 deg$^2$.
  Thus 
    one XRT tile
  would be $\delta A \sim$0.15 deg$^2$,
    or about ten times the minimum required to tile a putative short GRB host galaxy.
          Given an XRT exposure time of $\sim$0.1 ks (Kanner et al. 2012)
    and a comparable time to slew between tiles, a complete
    search of the three error boxes depicted in Figure 4
               would place modest demands on \emph{Swift}.
      The bright galaxy strategy for GW-EM follow-up
   described in this work
      results in a far less strenuous use of XRT resources
                           than discussed for example in Evans et al. (2015).

\subsection{Radio}

Explosive transients which eject ionized matter into a surrounding
   medium eventually produce radio waves  via synchrotron radiation
     as the electrons interact with tangled magnetic fields in shocked regions.
         Radio emission has been seen in GRBs, supernova remnants, colliding
     winds in massive binaries, symbiotic stars, and cataclysmic variables.
           Over long timescales 
 the radio provides  good calorimetry on the total energetics of the explosion
   due to its quasi-isotropic emission 
   (Nakar \& Piran 2011; Hotokezaka \& Piran 2015). Thus,
      as with the optical observations
         of kilonovae, the radio band also has
            an advantage over X-rays,  which are  beamed.
   The radio sky is also relatively quiet as regards fast transients,
         although this may be at least partly due to the lower sensitivity of
                transient radio surveys compared to optical or X-ray ones.
   (Frail et al. 2012; Mooley et al. 2013, 2015).

  Radio facilities have done extensive follow-up work on GRBs
      and would be relevant in radio tilings of GW error boxes
   insofar as having beam FWHMs that would 
      cover the expected region in an $L\simeq L^*$ galaxy:
 e.g., 
   AMI-LA: 6 arcmin beam at 16 GHz and
    JVLA:  9 arcmin beam in C band (4.5$-$5 GHz).
 At least 4 short GRBs have had radio afterglows detected, 
  051221a,
  050724,
  130603b, and
  140903 (Chandra \& Frail 2012; Berger 2014).   They were all quite faint, 
     $\sim$100 $\mu$Jy, but they were also considerably
       beyond the range of 
         current interest $d \le 200$ Mpc or $z \le 0.047$.

\section{Discussion and Conclusion}
 
 In this work we have argued for a strategy in which
    only the brighter galaxies are considered in GW-EM follow-ups.
    We show that the CLU catalog is fairly complete  
     for $x > x_{1/2}$ out to 200 Mpc.
           By weighting the galaxy counts within projected sky areas
           with a simple model for the aLIGO radial localization, 
    we find that only about $18\pm5$ (for 2015), $23\pm4$ (for 2017), or $11\pm2$ (for 2020)
   galaxies  need be considered, assuming error boxes of $500 $, $100 $ and $20$ deg$^2$,
        respectively. This results
             if one restricts attention
  to galaxies in the upper 50th percentile in integrated luminosity density.
      Furthermore, there are numerous EM detectors  
   with the property that one tile (i.e., one FOV) 
     would encompass the region of interest.
         These facilities -- optical, X-ray, and radio detectors --
      have carried out GRB follow-up previously  and could participate
                       in future  tiling observations.
   Having one tile per galaxy reduces the number of requisite
tiles down to the number of galaxies, a reduction in tiling
   effort by $\sim$$1-2$ orders of magnitude.

The CLU efficiencies for $x > x_{1/2}$ vary from $\sim$100\% at NS-NS inspiral range $\mu=60$ Mpc to 
$\sim$40\% at $\mu=180$ Mpc. For our third target year, 2020, the efficiency is
cut an additional factor of two due to the fact that $\sim$half of the detections 
     at any given time come from beyond the range, and the CLU is truncated beyond 200 Mpc.
       In addition to the CLU incompleteness, our adopted cut $x_{\rm cut} = x_{1/2}$ lies
   at roughly the median in the observed sGRB host galaxy $L_B/{L_B}^*$ distribution (Berger 2014),
         therefore we lose another factor of two in potential sGRB hosts. 
               Lowering our $x_{\rm cut}$ value
    further would allow us to cover more of the expected sGRB host galaxy $L_B/{L_B}^*$ distribution
       but would exacerbate the CLU incompleteness issue.
          An additional caveat arises from the hostless sGRBS with optical afterglows,
   and the (majority) of sGRBs for which no afterglows have been observed.  If these lie
     further from the host galaxy centers than the $\sim$100 kpc maximal offset
      indicated by  Fong \& Berger (2013), then our strategy could miss the kilonova emission.

M.M.K. acknowledges generous support from the Carnegie-Princeton Fellowship.
S.N. and L.P.S. thank the Aspen Center for Physics and the NSF Grant
      \#1066293 for hospitality during the editing of this paper.
S.N. acknowledges generous support from the Radboud University Excellence Initiative.
    We thank internal LIGO reviewer Ilya Mandel for excellent feedback on all aspects of the paper.

%\begin{thebibliography}{75}
%\expandafter\ifx\csname natexlab\endcsname\relax\def\natexlab#1{#1}\fi

\end{document}